\documentclass[11pt,preprint]{aastex}

\usepackage{graphics,graphicx}
\usepackage{natbib}
\newcommand{\gtorder}{\mathrel{\raise.3ex\hbox{$>$}\mkern-14mu
            \lower0.6ex\hbox{$\sim$}}}
\newcommand{\ltorder}{\mathrel{\raise.3ex\hbox{$<$}\mkern-14mu
            \lower0.6ex\hbox{$\sim$}}}

\shorttitle{An Upper Bound on Neutron Star Masses}
\shortauthors{Lawrence et al.}

\begin{document}

\title{AN UPPER BOUND ON NEUTRON STAR MASSES FROM MODELS OF SHORT GAMMA-RAY BURSTS}

\author{Scott Lawrence\altaffilmark{1}, Justin G. Tervala\altaffilmark{1}, Paulo F. Bedaque\altaffilmark{1}, M. Coleman Miller\altaffilmark{2}}

\affil{
{$^1$}{Department of Physics, University of Maryland, College Park, MD 20742-2421 USA; srl@umd.edu}
{$^2$}{Department of Astronomy and Joint Space-Science Institute, University of Maryland, College Park, MD 20742-2421 USA}
}

\begin{abstract}

The discovery of two neutron stars with gravitational masses $\approx 2~M_\odot$ has placed a strong lower limit on the maximum mass of nonrotating neutron stars, and with it a strong constraint on the properties of cold matter beyond nuclear density.  Current upper mass limits are much looser.  Here we note that, if most short gamma-ray bursts are produced by the coalescence of two neutron stars, and if the merger remnant collapses quickly, then the upper mass limit is constrained tightly.  If the rotation of the merger remnant is limited only by mass-shedding (which seems probable based on numerical studies), then the maximum gravitational mass of a nonrotating neutron star is $\approx 2-2.2~M_\odot$ if the masses of neutron stars that coalesce to produce gamma-ray bursts are in the range seen in Galactic double neutron star systems.  These limits would be increased by $\sim 4$\% in the probably unrealistic case that the remnants rotate at $\sim 30$\% below mass-shedding, and by $\sim 15$\% in the extreme case that the remnants do not rotate at all.  Future coincident detection of short gamma-ray bursts with gravitational waves will strengthen these arguments because they will produce tight bounds on the masses of the components for individual events.  If these limits are accurate then a reasonable fraction of double neutron star mergers might not produce gamma-ray bursts.  In that case, or in the case that many short bursts are produced instead by the mergers of neutron stars with black holes, the implied rate of gravitational wave detections will be increased.

\end{abstract}


\section{INTRODUCTION}
\label{sec:introduction}

The state of cold matter beyond nuclear density cannot be resolved strictly with laboratory experiments, and nuclear theories diverge strongly in their predictions for such matter.  Thus astronomical observations are sought for guidance.  Important constraints were obtained from the discovery of stars with gravitational masses of $M=1.97\pm 0.04~M_\odot$ (PSR~J1614-2230; \citealt{2010Natur.467.1081D}) and $M=2.01\pm 0.04~M_\odot$ (PSR~J0348+0432; \citealt{2013Sci...340..448A}).  Radius measurements would be helpful, but current estimates are dominated by systematic errors although hope exists for future X-ray and gravitational wave measurements (see \citealt{2013arXiv1312.0029M} for an extensive discussion).  It would also be useful to have an upper limit to the maximum gravitational masses of nonrotating stars, in addition to the current $M_{\rm max}\gtorder 2~M_\odot$ lower limit (note that PSR~J0348+0432 has a spin period of 39~ms [\citealt{2013Sci...340..448A}], which is long enough compared to the $\sim 1$~ms minimum that for our purposes this star rotates slowly).  However, the most rigorous existing upper limits are not very restrictive: \citet{1974PhRvL..32..324R} found $M_{\rm max}\ltorder 3.2~M_\odot$, and \citet{1996ApJ...470L..61K} performed an updated treatment that found $M_{\rm max}\ltorder 2.9~M_\odot$, where the differences depend primarily on the density up to which we believe we know the equation of state (EOS) of cold matter.

Ongoing observations of short gamma-ray bursts, and recent theoretical considerations of their mechanism, might provide a route to tighter upper limits on the maximum mass.  Short gamma-ray bursts have long been thought to be most probably caused by the merger of either two neutron stars or a neutron star and a black hole (see \citealt{2014ARA&A..52...43B} for a recent review).  From the standpoints of energetics and time scales other candidates exist, such as the accretion-induced collapse of a white dwarf followed by magnetar-like rapid spindown (e.g., \citealt{1992ApJ...388..164D,1997ApJ...482..383Y,2008MNRAS.385.1455M}).  However, in addition to the burst rate being roughly consistent with the expected compact object merger rate (e.g., \citealt{2012ApJ...756..189F,2015MNRAS.448.3026W}), and magnetar models being disfavored by the lack of confirmed radio transients \citep{2015arXiv150201350M}, the merger picture has received recent support from estimates of the spatial offsets of bursts from their most likely galactic hosts \citep{2006ApJ...648.1110B,2010ApJ...708....9F,2011MNRAS.413.2004C,2013ApJ...776...18F,2014ApJ...792..123B}.  These offsets are broadly consistent with expectations based on the supernova recoil that accompanies the births of neutron stars and black holes, and on the time needed after formation for the binary to coalesce due to the emission of gravitational waves.  

The compact object coalescence model for short gamma-ray bursts has been explored numerically and analytically with progressively greater fidelity over the last several years.  Observations of afterglow spectral changes characteristic of jet breaks imply that short bursts, like their long counterparts, have large bulk Lorentz factors $\Gamma>100$ \citep{1991ApJ...373..277K,1993A&AS...97...59F,1997ApJ...491..663B}.  Thus the jets that we see from short bursts must avoid being loaded with too many baryons, because this would slow down the jet \citep{1990ApJ...365L..55S} and would delay and lengthen the burst.  If the compact objects are a black hole and a neutron star, it is believed that this happens automatically because the orbital axis will naturally be nearly free of baryons due to the existence, from the beginning, of an event horizon in the system.

However, if the compact objects are both neutron stars then contamination of the jet by baryons is possible.  \citet{2014ApJ...788L...8M} argue that in order to produce the observed high Lorentz factors, the merger remnant must collapse within no more than 100 milliseconds after the initial merger, otherwise baryons driven outwards by interactions with neutrinos would overload the jet and cause the burst to last longer than is typical for short gamma-ray bursts (see also \citealt{2002MNRAS.336L...7R}).  In this sense, it is similar to the argument that the duration of a long gamma-ray burst is comparable to or larger than the time needed for the jet to escape the envelope of the massive star whose collapse caused the burst.  For an alternate view see the ``time-reversal" scenario of \citealt{2015ApJ...798L..36C}, who suggest that delayed collapse of a uniformly rotating star can explain both the prompt gamma-ray emission and the occasional $10^2-10^5$~s X-ray emission that follows; note, however, that \citealt{2015arXiv150501842M} argue that the collapse of a uniformly rotating star will not produce a disk and therefore will not generate a powerful jet.  More work is clearly required.

If this picture is correct and if most short gamma-ray bursts indeed come from mergers of two neutron stars rather than from the merger of a neutron star and a black hole, then we can infer an upper limit to the masses of nonrotating neutron stars.  The essence of the idea, which we elaborate in Section~\ref{sec:methods}, is that EOSs that predict large maximum masses for nonrotating stars also predict that the combined mass of two neutron stars can be supported stably in a uniformly rotating configuration, which would mean that short gamma-ray bursts are not produced\footnote{As we were performing the work for this paper we learned that a similar idea for limiting neutron star maximum masses, with somewhat different methodology, was being prepared by \citet{2015arXiv150407605F}.}.  Assuming that the double neutron star systems we see in our Galaxy are representative of the population that produces short gamma ray bursts, we show that this leads to tight constraints on the maximum mass.  We find that if the uniform rotation of merged remnants is limited only by mass-shedding (which is consistent with published neutron star merger simulations) then the maximum gravitational mass for nonrotating stars is $\approx 2-2.2~M_\odot$, depending on the masses of the double neutron stars that produce gamma-ray bursts and somewhat on the EOS class that we consider.  Although remnant rotation at the mass-shedding limit is consistent with all current simulations, we also consider more slowly rotating remnants, which are conceivable if a large fraction of the angular momentum is removed from the system very soon after merger.  In the extreme case that the remnant has no angular momentum our upper mass limits are increased by $\sim 15$\%.

These tight limits would have important implications for cold matter beyond nuclear density.  Depending on the range of masses of neutron stars in mergers, this could also imply that a fair fraction of mergers do {\it not} lead to short gamma-ray bursts.  This raises the question of what form their still-large energy release would take.  It also implies that the merger rates inferred from short gamma-ray bursts need to be increased.  This would increase the likely merger detection rates for ground-based gravitational wave detectors such as Advanced {\it LIGO}, Advanced Virgo, and {\it KAGRA} \citep{2014arXiv1411.6068D}.

If instead short gamma-ray bursts are produced by the coalescence of a neutron star with a black hole, then our argument does not apply and we therefore cannot use this argument to place an upper limit on the maximum mass of a nonrotating neutron star.  This is because, in that case, an event horizon already exists and the jet funnel will therefore presumably already be clean.  The news would then be even better for ground-based gravitational wave detectors, because given that black holes are more massive than neutron stars, the signal produced will be stronger and thus visible to greater distances than the signal from a double neutron star merger.  Therefore, for a given rate of short gamma-ray bursts per volume, the detection rate will be significantly greater than for double neutron star mergers.

In Section~\ref{sec:methods} we discuss our assumptions and method of analysis.  In Section~\ref{sec:results} we give our results.  We motivate the parametrized forms of our EOSs and the allowed ranges for those parameters, and then give the mass limits that result.  We finish in Section~\ref{sec:summary} by discussing how future electromagnetic observations of gamma-ray bursts that are also detected with gravitational wave instrumentation can make our results more rigorous.  In particular, we show that even if only the chirp mass, rather than both masses independently, can be measured, the uncertainty about the separate masses has only a small effect on our constraints.

\section{METHODS}
\label{sec:methods}

Our primary tool for analysis is the publicly available Rotating Neutron Star ({\tt rns}) code developed by \citet{1995ApJ...444..306S}.  This is a flexible code that computes the structure and external spacetime of an axisymmetric, uniformly rotating star for a given EOS.  We have modified the code slightly so that for a given EOS it can output (1)~the maximum gravitational mass for a nonrotating star, (2)~the baryonic rest mass corresponding to a given gravitational mass for a nonrotating star, and (3)~the maximum baryonic rest mass for a uniformly rotating configuration, which is limited by either mass-shedding alone or, in addition, by a limit on the ratio $T/W$ of the rotational kinetic energy to the gravitational binding energy (note that $T/W$ at mass shedding typically ranges from $\sim 0.1$ for very soft EOS to $\sim 0.15$ for very stiff EOS; see, e.g., Table~5 of \citealt{1994ApJ...424..823C}).  One way in which a threshold on $T/W$ could be relevant is if the merged remnant develops a nonaxisymmetric instability, because in principle such nonaxisymmetries could lead to the emission of gravitational waves that would reduce $T/W$ to the threshold value. However, treatments of relativistic fluids suggest that the threshold value is $T/W>0.2$ for realistic neutron star compactnesses \citep{1998ApJS..117..531S,2002PhRvD..66d4021G}, which is well above the mass-shedding limit.  Thus secular instabilities are not likely to be important, so we consider $T/W$ thresholds only as a way to determine the effect on our limits of, e.g., rapid angular momentum loss due to magnetic braking.

We have tested our implementation of the code by comparing our answers to (1)--(3) above with those given in \citet{1994ApJ...424..823C} for several tabulated EOSs and in \citet{2009PhRvD..79l4032R} for several parametrized approximations to other EOSs.  We find agreement to better than about 1\% in all cases except for the very soft EOSs F and G used in \citet{1994ApJ...424..823C}; these EOSs are ruled out by the existence of neutron stars with gravitational masses $\approx 2~M_\odot$.  We note that for most of the EOSs listed in \citet{1994ApJ...424..823C}, the sound speed becomes superluminal ($c_s^2=dP/d\rho>c^2$) above some energy density that is usually less than the central energy density for the maximum mass rotating and nonrotating configurations.  If we impose a $c_s=c$ upper limit then the maximum masses often decrease significantly: for example, for EOS UU, the maximum gravitational mass for a nonrotating star drops from $M_{\rm max}=2.20~M_\odot$ to $2.06~M_\odot$ and the maximum baryonic rest mass for a uniformly rotating star drops from $M_{\rm bary,rot,max}=3.12~M_\odot$ to $2.94~M_\odot$.

Our limits are conservative in the sense that we assume that the remnant is cold, and that it rotates uniformly rather than differentially.  A hot remnant can support more mass than a cold remnant (see, e.g., the recent treatment in \citealt{2014ApJ...790...19K}), and differential rotation can support more mass than uniform rotation (e.g., \citealt{2000ApJ...528L..29B}).  However, we expect that temperatures comparable to the Fermi temperature $\sim 10^{12}$~K, which are required to provide significant extra support, will exist for at most an extremely short time because neutrino emission will remove the energy efficiently.  We also expect, as suggested by \citet{2000ApJ...544..397S}, that internal magnetic fields will be amplified rapidly enough by differential motion that the angular momentum will be redistributed into a state of uniform rotation.  If either of these assumptions is incorrect then the upper limit to neutron star masses will be lowered somewhat.  Our upper limit would also be lowered if some of the baryonic mass in the two stars ends up in a disk or outflow rather than as part of the merged remnant, although in current merger simulations the escaping mass typically amounts to only ${\rm few}\times 10^{-2}~M_\odot$ (e.g., \citealt{2000PhRvD..61f4001S,2003PhRvD..68h4020S,2004PhRvD..69j4030D,2008PhRvD..78b4012L,2009PhRvD..80f4037K,2010CQGra..27k4105R,2011ApJ...739...47F,2013ApJ...762L..18G,2014PhRvD..89j4021B,2015PhRvD..91f4027K,2015arXiv150401266D}).

Our assumptions might not be conservative if the dynamics of neutron star mergers are such that the remnant rotates at a rate less than the mass-shedding limit for the remnant.  Current simulations (e.g., \citealt{2014PhRvD..89j4021B,2015PhRvD..91f4027K,2015arXiv150401266D}) suggest that the total angular momentum of the remnant plus disk and outflows does exceed the mass-shedding limit, so we consider this our standard case.  However, if the rotation limit is tighter, less mass can be supported by the spinning remnant than in our assumptions, so a larger range of EOSs would satisfy our criterion for short gamma-ray bursts, and hence the maximum allowed gravitational mass for a nonrotating star would be larger than the mass we derive in our standard analysis.  Our mass limit would also be increased if the remnant has a strong enough poloidal magnetic field (either initially, or developed via, e.g., a dynamo induced by differential rotation) that it spins down significantly during the $\sim 0.1$~s allowed by the argument of \citet{2014ApJ...788L...8M}.  This would require a poloidal field with an extremely large characteristic surface strength of at least $B\sim {\rm few}\times 10^{16}$~G.  To take these possibilities into account we also explore the mass limits that come from the assumption that (as an arbitrary, round number) $T/W=0.1$ limits the rotation, and the mass limits that would apply in the extreme case that the remnant is nonrotating.

Our method is the following:

\begin{enumerate}

\item Start with an assumed EOS and two neutron stars, which we assume to be nonrotating (because the fastest-spinning neutron star in a double neutron star system has a frequency of only 44~Hz [\citealt{2003Natur.426..531B}] and tidal torques will not spin the stars up significantly even near merger [\citealt{1992ApJ...400..175B,1992ApJ...398..234K}]).  Let the gravitational masses be $M_1$ and $M_2$, and let their corresponding baryonic rest masses for the chosen EOS be $M_{\rm bary,1}$ and $M_{\rm bary,2}$.  We investigate three pairs of masses: $M_1=1.25~M_\odot$ and $M_2=1.35~M_\odot$ (comparable to PSR~J0737-3039A,B, which is the the lightest double neutron star pair yet discovered [\citealt{2003Natur.426..531B,2004Sci...303.1153L}]); $M_1=M_2=1.35~M_\odot$ (comparable to the average gravitational mass of double neutron stars discovered in our Galaxy, and similar to the systems PSR~B1534+12 [\citealt{2002ApJ...581..501S}] and PSR~B2127+11C [\citealt{2006ApJ...644L.113J}]); and $M_1=1.35~M_\odot$ and $M_2=1.45~M_\odot$ (comparable to PSR B1913+16, which is the heaviest double neutron star pair yet discovered [\citealt{1992RSPTA.341..117T,2010ApJ...722.1030W}]).  See \citet{2013ApJ...778...66K} for a recent summary of the masses in double neutron star systems.

\item If, when they merge, the stars produce a gamma-ray burst, then by the logic of \citet{2014ApJ...788L...8M} the remnant must collapse quickly to form a black hole.  Thus the baryonic rest mass $M_{\rm rem,bary}$ of the remnant must exceed the stable limit of a uniformly rotating neutron star.

\item We assume that $M_{\rm rem,bary}=M_{\rm bary,1}+M_{\rm bary,2}$, which is the maximum possible.  Any matter that goes into an outflow or a disk that lasts for more than 0.1~s will reduce the remnant mass and strengthen our argument.

\item We compare $M_{\rm rem,bary}$ with the maximum baryonic rest mass $M_{\rm bary,rot,max}$ that can be sustained by a uniformly rotating star for the assumed EOS.  The rotation is limited by either (a)~mass-shedding, or (b)~a limit on $T/W$, whichever is more restrictive.  Our primary results are based on (a), which we believe to be the most realistic case, but we also explore $T/W=0.1$ and $T/W=0$ (nonrotating).  

\item If $M_{\rm bary,rot,max}<M_{\rm rem,bary}$ then the remnant collapses and the equation of state is viable.  For this equation of state we can also compute the maximum gravitational mass $M_{\rm max}$ for a nonrotating star.

\item We therefore search the parameter space in our EOSs to find the largest $M_{\rm max}$ that is viable by our short gamma-ray burst criterion.  This is the number we report.

\end{enumerate}

\section{RESULTS}
\label{sec:results}

\subsection{Parametrized equations of state}

The neutron Fermi momentum in pure neutron matter at densities below twice nuclear saturation density ($n<2n_s$, where $n_s=1.6\times 10^{38}~{\rm cm}^{-3}$; the corresponding mass density is $\rho_s=m_nn_s=2.7\times 10^{14}~{\rm g~cm}^{-3}$) is less than $\sim 420$~MeV/c. At $n=2n_s$, two neutrons with opposite momenta on the top of the Fermi sphere have a total energy of about $320$ MeV (in the lab frame) or $160$ MeV (in the center of mass frame). This is barely above the pion production threshold of $\approx 140~{\rm MeV}$ and indicates that a treatment of dense neutron matter based on neutron degrees of freedom interacting through a potential should be adequate. The character of interactions between the nucleons can be inferred from elastic nucleon-nucleon scattering data and the spectroscopy of light nuclei ($A\leq 20$). Modern many-body methods, whether computational or analytical, can then be used to infer the zero temperature EOS. This program has been carried out in the last few years and represents an important step towards an understanding of dense matter from first principles, at least at the low end of densities relevant to neutron stars.

There are two versions in the literature of the program sketched above. The first one \citep{2010PhRvL.105p1102H,2010PhRvC..82a4314H} attempts to describe the nucleon-nucleon interaction using effective field theory ideas \citep{1991NuPhB.363....3W,2002ARNPS..52..339B,2000nucl.th...8064B,2002CzJPh..52B..49P,2009RvMP...81.1773E} to extract the nucleon-nucleon potential from quantum chromodynamics. In this approach there is a systematic expansion of the interaction in powers of the momentum and pion masses and uncertainties can be quantified a priori. Three-body forces appear as a small but important effect. Unfortunately, this approach describes the nucleon-nucleon phase shifts up to relatively small momentum and is adequate only for densities below nuclear saturation. 

A second approach is more useful for our purposes \citep{2012PhRvC..85c2801G,2014EPJA...50...10G}, and it forms the basis for our primary parametrized equation of state, which we call EOS1.  In this approach, phenomenological nucleon-nucleon potentials are fit to nucleon-nucleon elastic scattering data.  These potentials describe the phase shifts well up to energies around $600$ MeV (in the lab frame). Two-nucleon potentials are, however, not enough to describe matter even below nuclear saturation density.  Effective theory arguments as well as studies of light nuclei demonstrate that three-body forces are required. The importance of the three-body forces increases with density and is substantial at $n = 2 n_s$. Some components of the three-body force can be extracted by fitting light nucleus energy levels. However, in neutron matter, unlike in nearly symmetric matter, only the isospin-3/2 channel is relevant, so the three-body components that are derived from light nuclei are not the dominant ones in neutron matter.  Instead, \citet{2012PhRvC..85c2801G,2014EPJA...50...10G} used a variety of three-body forces with differing functional forms and ranges varying over a factor two around the pion Compton wavelength, where the strength of each force was fixed so that the symmetry energy lies within the empirically observed range \citep{2012PhRvC..86a5803T}.  These interactions were used to obtain the EOS  using the auxiliary field Green's function Monte Carlo method \citep{1999PhLB..446...99S}.  The errors arising from the Monte Carlo and infinite volume extrapolation are negligible for our purposes.  The three-body force is, then, the largest source of uncertainty about the neutron matter EOS at densities below $2 n_s$. 

The results of \citet{2012PhRvC..85c2801G,2014EPJA...50...10G} are well fit with the convenient parametrization
\begin{equation}
\epsilon(n) = n \left[m_n + a\left(n/n_s\right)^\alpha +b\left(
n/n_s\right)^\beta\right]
\label{eq:EOS1}
\end{equation} 
for the mass-energy density $\epsilon$ as a function of the number density $n$.
The parameters $a$ and $\alpha$ depend primarily on the two-body force and are well constrained by scattering data to lie in the ranges $12.6~{\rm MeV} < a < 13.0~{\rm MeV}$ and $0.47 < \alpha < 0.50$. The parameters $b$ and $\beta$ are more sensitive to the three-body force; in the analysis of \citet{2012PhRvC..85c2801G,2014EPJA...50...10G} they vary over the ranges $3.2~{\rm MeV}< b < 5.2~{\rm MeV}$ and $2.1 < \beta < 2.5$ \footnote{There is a strong correlation between the parameters $b$ and $\beta$ that further constrains the set of equations of state. In the spirit of being as conservative as possible we will neglect this correlation in our work.}. However, in order to be conservative, we will present results obtained by doubling the uncertainty in these parameters.  Thus the ranges we search are $12.4~{\rm MeV}<a<13.2~{\rm MeV}$, $0.45<\alpha<0.52$, $2.2~{\rm MeV}<b<6.2~{\rm MeV}$, and $1.9<\beta<2.7$.

For EOS1 we adopt the equation of state given above to a threshold rest mass density $\rho_{\rm thresh,1}$, where we explore the range $1.7~\rho_s<\rho_{\rm thresh,1}<2~\rho_s$.  Between $\rho_{\rm thresh,1}$ and $\rho_{\rm thresh,2}$ (which we do not constrain except to require that it exceed $\rho_{\rm thresh,1}$) we assume a constant sound speed $c_s$ that we allow to be anywhere between $c_s=c/2$ and $c_s=c$ (the low-density equation of state in Equation~\ref{eq:EOS1} is itself never acausal).  We assume that $c_s=c$ above $\rho_{\rm thresh,2}$ because the central mass-energy density of a maximum mass nonrotating star is greater than that of a maximum mass uniformly rotating star (compare, e.g., the central densities of the nonrotating stars in Table 4 of \citealt{1994ApJ...424..823C} with the central densities of the rotating stars in Table 5 of \citealt{1994ApJ...424..823C}).  Therefore a transition above some density to the hardest possible EOS can increase the maximum mass of a nonrotating star without affecting the maximum mass of a uniformly rotating star.

The class of equations of state included in EOS1 is very large and was designed with the objective of maximizing the maximum non-rotating mass while allowing for short gamma-ray bursts to follow mergers. Still, it does not include some possibilities that, while not favored, are not excluded by hard evidence.  The calculations leading to the low density part of the EOS1 assume that nucleons (and electrons and muons) are the only relevant degrees of freedom at densities below $1.7 n_s$. This expectation can be frustrated if a pion condensate \citep{1978RvMP...50..107M}, or hyperons, become important at these low densities. Equations of state with any extra degrees of freedom besides nucleons tend to be too soft to support a maximum non-rotating mass of $2 M_\odot$ but the possibility remains that the correct equation of state is softer than EOS1 at $n<n_s$ but stiffens quickly at higher densities. A similar situation would also be obtained in the even more unlikely possibility that a transition to quark matter occurs at densities below $1.7~n_s$ \citep{1999LNP...516..162M,2008RvMP...80.1455A}. Finally, a very different scenario arises if strange quark matter is the true ground state of matter at arbitrarily low densities. In this case the low density equation of state would have a very different form from EOS1 \citep{1971PhRvD...4.1601B,1984PhRvD.30..272W} and our results would not apply at all.

Our secondary parametrized EOS, which we call EOS2, is a slight modification of the piecewise polytrope introduced by \citet{2009PhRvD..79l4032R}.  EOS2 represents the pressure as a function of density with four parameters $\rho_0$, $\Gamma_1$, $\Gamma_2$, and $\Gamma_3$:
\begin{equation}
\begin{array}{rll}
P&\propto \rho^{1.58425}&\rho<2.44034\times 10^7~{\rm g~cm}^{-3}\\
&\propto \rho^{1.28733}&2.44034\times 10^7~{\rm g~cm}^{-3}<\rho<3.78358\times 10^{11}~{\rm g~cm}^{-3}\\
&\propto \rho^{0.62223}&3.78358\times 10^{11}~{\rm g~cm}^{-3}<\rho<2.6278\times 10^{12}~{\rm g~cm}^{-3}\\
&=3.594\times 10^{13}~{\rm dyn~cm}^{-2}~\rho^{1.35692}&2.6278\times 10^{12}~{\rm g~cm}^{-3}<\rho<\rho_0\\
&\propto \rho^{\Gamma_1}&\rho_0<\rho<10^{14.7}~{\rm g~cm}^{-3}\\
&\propto \rho^{\Gamma_2}&10^{14.7}~{\rm g~cm}^{-3}<\rho<10^{15}~{\rm g~cm}^{-3}\\
&\propto \rho^{\Gamma_3}&10^{15}~{\rm g~cm}^{-3}<\rho\; .\\
\end{array}
\end{equation}
The normalization in the $2.6278\times 10^{12}~{\rm g~cm}^{-3}<\rho<\rho_0$ range is taken from {\bf Table~II of \citet{2009PhRvD..79l4032R}; note that their table lists the normalization divided by $c^2$.}  

We impose the additional limitation that if there is any density $\rho_{\rm caus}$ at which these expressions would predict that the EOS becomes acausal above some density $\rho_{\rm caus}$, we set $c_s=c$ at $\rho>\rho_{\rm caus}$ (see \citealt{2007GReGr..39.1651E} for a discussion of the causality limit).  That is, we do not constrain the parameter space $(\rho_0,\Gamma_1,\Gamma_2,\Gamma_3)$ based on the requirement of causality; instead, if a given combination predicts $c_s>c$ above some density, we set the sound speed to be equal to $c$ above that density.  {\bf This approach generalizes somewhat the EOS of \citet{2009PhRvD..79l4032R}, and is consistent with our philosophy of considering as broad a set of EOSs as possible.}  The prefactors for the pressure in each density range are set by pressure continuity at each of the density boundaries, which are anchored by the pressure in the $2.6278\times 10^{12}~{\rm g~cm}^{-3}<\rho<\rho_0$ range, and all the densities are measured in g~cm$^{-3}$.  The ranges that we search in the parameters are $\rho_0=(1-8)\times 10^{-4}M_\odot/(GM_\odot/c^2)^3=(1-8)\times 6.173\times 10^{13}~{\rm g~cm}^{-3}$; $\Gamma_1=1.5-5$; $\Gamma_2=1-5$; and $\Gamma_3=1-5$. 

\subsection{Mass limits}

We use Powell's direction set method \citep{1964CompJ...7..155P} to maximize the maximum gravitational mass of a nonrotating star over the parameters in each of our EOS models.  For both EOSs, the maximum mass is nearly unimodal over the parameter space, which makes searches relatively smooth and reproducible.  

\begin{figure}[!htb]
\begin{center}
\plotone{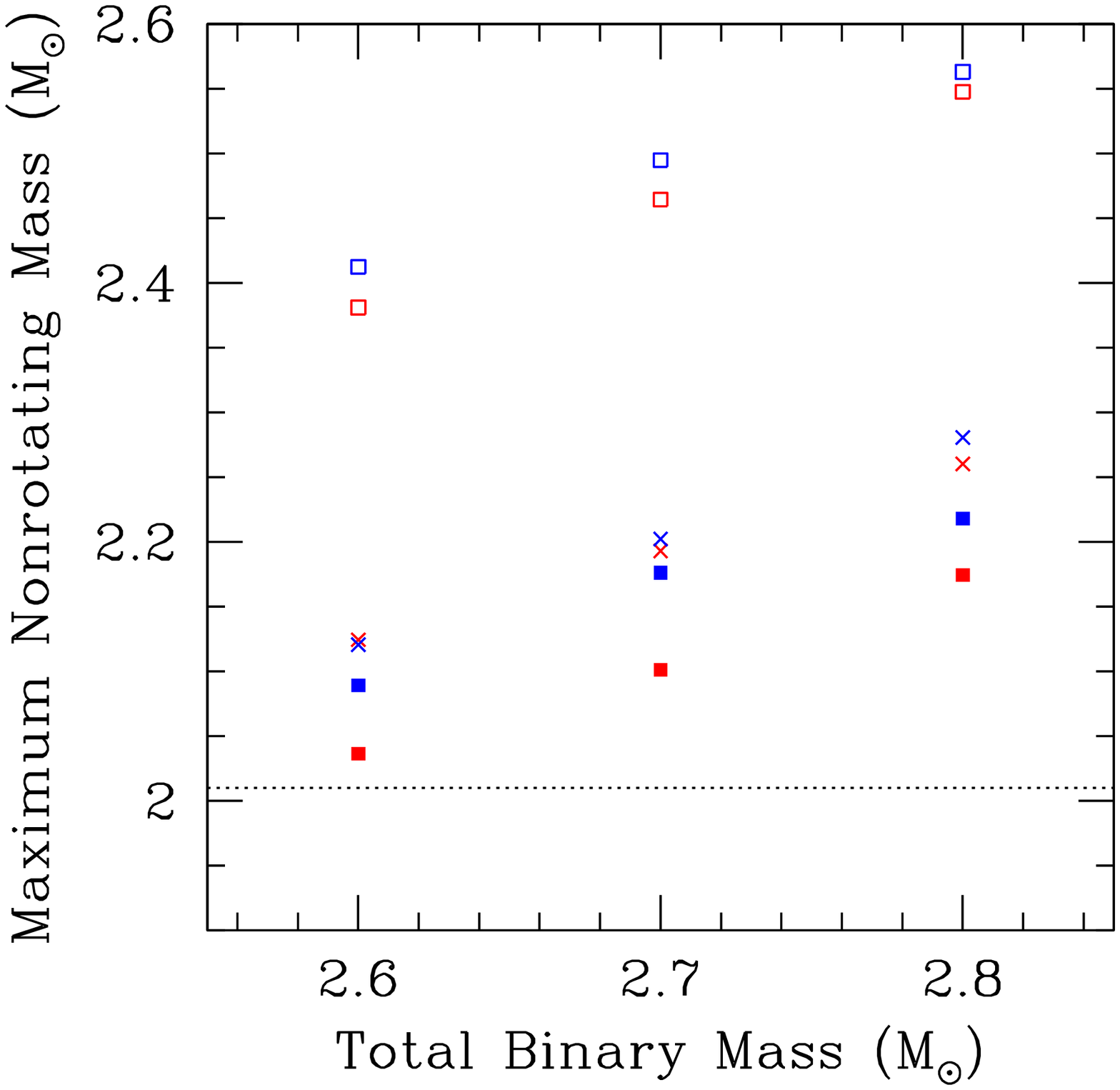}
\vskip-1.5cm
\caption{Upper limits on the maximum gravitational mass of a nonrotating star, based on the criterion that to make gamma-ray bursts, double neutron star mergers must produce an object that collapses when it becomes uniformly rotating.  Red symbols refer to EOS1 and blue symbols refer to EOS2 (see text for descriptions of these EOSs).  The three types of mergers we consider are of gravitational masses $1.25~M_\odot$ and $1.35~M_\odot$; of $1.35~M_\odot$ and $1.35~M_\odot$; and of $1.35~M_\odot$ and $1.45~M_\odot$, to represent the range of masses in known double neutron star binaries.  Filled squares show the limits obtained when we assume that the rotation of the merged remnant is limited only by mass shedding, which is our standard case and which is consistent with existing simulations.  To explore the probably unrealistic situation in which there is very rapid angular momentum loss from the system, we also consider cases in which there are stricter angular momentum limits: crosses show the masses obtained when we assume that the more restrictive of mass shedding or $T/W<0.1$ limit the rotation of the remnant, and the open symbols are for nonrotating remnants.  The dotted horizontal line shows the maximum currently known gravitational mass $2.01~M_\odot$ for a neutron star.  This figure demonstrates that if short gamma-ray bursts are produced by the merger of double neutron star systems comparable to the ones we see in our Galaxy, then the upper mass limit for nonrotating neutron stars is constrained tightly.
}
\label{fig:masslim}
\end{center}
\end{figure}

We display our results in Figure~\ref{fig:masslim}.  For each of the three combinations of gravitational mass $(1.25~M_\odot,1.35~M_\odot)$, $(1.35~M_\odot,1.35~M_\odot)$, and $(1.35~M_\odot,1.45~M_\odot)$ we show the mass limits we obtain for EOS1 and EOS2 and for different rotation limits: mass shedding only (no $T/W$ limit); the more restrictive of mass shedding and $T/W<0.1$; and a nonrotating remnant.  This figure shows that if short gamma-ray bursts are produced by mergers of double neutron stars similar to those we have discovered in the Galaxy and if the remnant is unstable in a uniformly rotating state, then the limits on the maximum mass of a nonrotating neutron star are extremely tight, particularly if mass shedding alone sets the limit on uniformly rotating merged remnants.  We see that the limits from EOS1 are tighter than the limits from EOS2, because EOS2 allows greater freedom for the EOS at high densities.  We also see that if low-mass double neutron star mergers produce short GRBs then the maximum gravitational mass of a nonrotating neutron star is already known with high precision.

\section{IMPLICATIONS AND SUMMARY}
\label{sec:summary}

We have shown that if short gamma-ray bursts are produced when mergers of two neutron stars lead to a rapid collapse to a black hole, and if the neutron star masses are similar to what we see in Galactic double neutron star systems, then the maximum mass of nonrotating neutron stars is constrained tightly from above as well as from below.  This is because if the maximum mass exceeds our limit, the merger of two neutron stars of the specified masses will produce an object that remains stable even after internal angular momentum transport produces uniform rotation.  Thus prompt collapse would not happen; any collapse would be delayed by at least as long as it would take to radiate angular momentum from the system.

Much more certain limits on the maximum mass will be obtained when ground-based gravitational wave detectors see the gravitational radiation from double neutron star coalescences in coincidence with short gamma-ray bursts.  This is because such coincidences will allow us to limit strongly the total baryonic rest mass of the merger.  The smallest such total mass associated with a successful burst will place the tightest upper limit on the maximum mass of a nonrotating neutron star. 

To determine how tightly we could constrain the total baryonic rest mass in a merger, we note that the mass-related quantity that will be most precisely measured from gravitational wave trains is the chirp mass
\begin{equation}
M_{\rm ch}\equiv \eta^{3/5}M_{\rm tot}
\end{equation}
where $M_{\rm tot}\equiv M_1+M_2$ is the total gravitational mass and $\eta\equiv M_1M_2/M_{\rm tot}^2=q/(1+q)^2$ is the symmetric mass ratio, where $q\equiv M_1/M_2\leq 1$ is the standard mass ratio.  The chirp mass is essentially estimated by counting gravitational wave cycles, so for a double neutron star coalescence it will be estimated to a precision that is typically better than $\sim 0.1$\% given that $\sim 10^3$ cycles are expected in the band of the detectors (see \citealt{2014arXiv1411.6934B} for a recent study).  Thus we can assume that $M_{\rm ch}$ will be determined exactly, but we cannot necessarily assume that the signal will be strong enough to break the degeneracy and determine both masses independently.  We can invert the equations to find the individual gravitational masses, given the unknown mass ratio $q$:
\begin{equation}
\begin{array}{rl}
M_1&=q^{2/5}(1+q)^{1/5}M_{\rm ch}\\
M_2&=q^{-3/5}(1+q)^{1/5}M_{\rm ch}\; .\\
\end{array}
\end{equation}
Therefore, if we infer a given chirp mass, there will be some uncertainty in the individual gravitational masses and thus in the total baryonic rest mass in the merger.  

This uncertainty, however, is only a few hundredths of a solar mass, at least for masses similar to what we see in double neutron star systems in our Galaxy.  This is fundamentally because neutron star masses are limited from below; for precisely measured masses the current lower limit is $1.25~M_\odot$, but for the sake of argument let us say that the true limit is $1.2~M_\odot$.  Therefore, for a given chirp mass, there is not much room to change the total baryonic mass.  As a specific example, suppose we consider a merger between stars with gravitational masses $M_1=1.35~M_\odot$ and $M_2=1.45~M_\odot$, in the context of the parameter combination for EOS1 that maximizes the gravitational mass of a nonrotating star for successful production of short gamma-ray bursts with these masses.  The chirp mass for this combination is $M_{\rm ch}=1.218~M_\odot$, and the total baryonic rest mass for the two stars combined and this EOS is $M_{\rm bary,tot}=3.078~M_\odot$.  If we take the extreme that one of the stars actually has a gravitational mass of $M_1=1.2~M_\odot$, then to keep the same chirp mass it is necessary that $M_2=1.64~M_\odot$.  The total baryonic rest mass for this combination is then $M_{\rm bary,tot}=3.136~M_\odot$.  Thus even for this extreme case the difference is less than $0.06~M_\odot$.  If the lower limit to the gravitational mass of a neutron star is $1.25~M_\odot$ instead of $1.2~M_\odot$ then the maximum total baryonic mass for this case drops to $3.107~M_\odot$.  If the chirp mass is smaller (as it will be if the gravitational masses of the neutron stars are lower) then the correction will be even less.  We conclude that gravitational wave measurement of just the chirp mass from a coalescence coincident with a gamma-ray burst will place strong constraints on the total baryonic rest mass for a given EOS.

If the maximum mass of a nonrotating neutron star is towards the high end of what we infer (say, $2.2~M_\odot$) then mergers between lower-mass neutron stars will produce a remnant that will not collapse quickly unless angular momentum is actually removed from the system rather than redistributed.  As we discussed, this likely requires the production of a very strong poloidal magnetic field within tens of milliseconds.  If this does not happen in most cases, then many mergers could fail to produce short gamma-ray bursts; it would be interesting to know the observed properties of such an event.  It would also suggest that estimates of the gravitational wave detection rate of double neutron star mergers based on the short gamma-ray burst event rate are conservative, because only some fraction of coalescences lead to bursts.

Another possibility is that many short gamma-ray bursts are actually produced by the coalescence of neutron stars with black holes rather than neutron stars with neutron stars.  The larger chirp masses of such events means they will be detectable to greater distances than double neutron star mergers.  Thus for a given observed gamma-ray burst rate per volume, the gravitational wave detection rate would be increased by a factor of several.

It is important to note that, although in the context of our parametrizations an EOS that predicts a mass in excess of our maximum would not lead to short gamma-ray bursts, the converse is not necessarily true: a maximum nonrotating mass below our limit does {\it not} guarantee burst viability.  For example, consider a $1.35~M_\odot-1.45~M_\odot$ merger, and let us use EOS1 with $a=13.3$~MeV, $\alpha=0.51$, $b=4.1$~MeV, $\beta=2.3$, $\rho_{\rm thresh,1}=1.85\rho_s$, $c_s=0.75c$, and $\rho_{\rm thresh,2}=3\rho_s$, but (unlike in our standard parametrization) we set $c_s=c/2$ above $\rho_{\rm thresh,2}$, then the maximum gravitational mass of a nonrotating star is $M_{\rm max}=2.06~M_\odot$.  This is well below our threshold for $1.35~M_\odot - 1.45~M_\odot$ mergers with remnants whose spin is limited only by mass shedding.  However, the total baryonic rest mass of the stars, $3.075~M_\odot$, is less than the maximum $3.108~M_\odot$ that could be supported by a uniformly rotating star, and therefore a short burst would not occur with this combination.  Thus individual EOSs should be tested against the short gamma-ray burst criterion using a code such as {\tt rns} \citep{1995ApJ...444..306S}.

Finally, we note that we find tighter constraints $M_{\rm max}\approx 2-2.2~M_\odot$ than are reported in the study of \citet{2015arXiv150407605F}, who give limits of $2.3-2.4~M_\odot$.  We believe that the primary reason for the difference is that \citet{2015arXiv150407605F} are concerned with the question of whether {\it any} significant fraction of double neutron star mergers will produce short gamma-ray bursts.  Thus they concentrate on the high end of the neutron star masses that emerge from their population synthesis calculations, which means that they find larger upper limits than we find by focusing on three mass categories that we know exist in the Galaxy.  

Again, we emphasize that when ground-based gravitational wave detectors see bursts with mergers, the lowest mass example of a successful burst will set the strongest limits on the maximum mass of a nonrotating neutron star.  To that end it will be helpful to know whether there are any electromagnetic signatures of rapid collapse that can be identified clearly and seen from a broader range of directions than the burst itself, because this would increase the otherwise small fraction of mergers detected using gravitational waves that can be evaluated using our argument.  For example, it has been proposed \citep{2014MNRAS.441.3444M,2015MNRAS.450.1777K} that prompt formation of a black hole will lead to nearly isotropic red emission from disk winds within several days, whereas delayed formation of a hole will produce bluer emission within roughly a day.  It is also conceivable that there is a signature in the gravitational wave emission itself of prompt collapse, although this is likely to be at frequencies $>2$~kHz where ground-based detectors are not especially sensitive.

In summary, we show that current models of short gamma-ray bursts involving double neutron star mergers imply a strong upper limit to the maximum mass of nonrotating neutron stars.  If some of these bursts come from stars towards the low mass end of what we see in our Galaxy, then with the parametrized equations of state we have explored the limit could be close to $M_{\rm max}\approx 2~M_\odot$.  If the bursts only come from neutron star binaries near the high end of our Galactic sample then the maximum could be $M_{\rm max}\approx 2.2~M_\odot$, but in that case short bursts are likely to occur only for high-mass neutron star binaries, which means that gravitational wave detection rates inferred from burst rates need to be increased.  In either case, direct detection of gravitational waves from neutron star binaries along with coincident gamma-ray bursts will constrain strongly the properties of cold matter beyond nuclear density.

\acknowledgements

We thank Chris Belczynski, Edo Berger, Fred Lamb, Ilya Mandel, Brian Metzger, and Enrico Ramirez-Ruiz for helpful discussions.  The numerical computations reported in this paper were carried out on the Deepthought cluster at the University of Maryland.  PFB acknowledges support from the U.S. Dept. of Energy through grant number DEFG02-93ER-40762.

\bibliography{masslim}

\end{document}